\documentclass[letterpaper, 10 pt, conference]{ieeeconf} 
\usepackage{graphicx}
\IEEEoverridecommandlockouts         
\usepackage{xcolor}
\usepackage{float}
\let\labelindent\relax
\usepackage{enumitem}
\setlist[enumerate]{leftmargin=*}
\newtheorem{assumption}{Assumption}
\newtheorem{definition}{Definition}

\newtheorem{theorem}{Theorem}
\newtheorem{remark}{Remark}
\usepackage{amsmath}
\usepackage{amssymb}
\usepackage{mathtools}
\newcommand{\RR}{\mathbb{R}}
\newcommand{\K}{\mathcal{K}}
\newcommand{\fav}{f_{\text{av}}}
\def\e{\varepsilon}

\title{\LARGE \bf
Stability of Slow-Fast Nonlinear Dynamics: Moving Equilibrium Case}

\author{G. Q. Bao Tran, Daniel Liberzon, and Hyungbo Shim
\thanks{G.~Q.~B.~Tran and D.~Liberzon are with Coordinated Science Laboratory, University of Illinois Urbana-Champaign, Urbana, IL 61801, USA (e-mail: {\tt\small \{baotran,liberzon\}@illinois.edu}). Their work is supported by the AFOSR MURI FA9550-23-1-0337 and NSF CMMI-2452534 grants. H.~Shim is with ASRI, Department of Electrical and Computer Engineering, Seoul National University, Seoul, South Korea (e-mail: {\tt\small hshim@snu.ac.kr}). His work is supported in part by the National Research Foundation of Korea (NRF) grant funded by the Korea government (MSIT) (No. RS-2022-00165417).}
}

\begin{document}

\maketitle
\thispagestyle{empty}
\pagestyle{empty}

\begin{abstract}
We study the stability of nonlinear systems subject to both slow and fast time variations. Both can have discontinuities, covering switched systems as a special case. The fast variation is assumed to be periodic; thus, we
rely on averaging to construct an average system. Importantly, the equilibrium of the average system depends on the slow variation and is assumed to be exponentially stable when this slow input is frozen. Using perturbation and Lyapunov analyses, we establish a practical stability result showing that the system state remains within a neighborhood of the moving equilibrium when the total variation (flows and jumps) of the slow variation is appropriately bounded and the fast input varies sufficiently fast. The result is illustrated via a nonlinear switched system with slow-fast switching and a mode-dependent equilibrium.
\end{abstract}

\section{INTRODUCTION}
\label{intro}
Consider a nonlinear system
\begin{equation}\label{eq:sys}
\dot{x}(t) = f(x(t),u_s(t),u_f(t/\e)),
\end{equation}
where $x \in \RR^n$ is the state, $\e > 0$ is a small parameter, and $u_s \in \RR^m$ and $u_f \in \RR^l$ are slowly- and fast-varying signals modeling time variations in the dynamics. We are interested in establishing sufficient conditions for the stability of system~\eqref{eq:sys}. When the fast variation is absent and the dynamics are linear, textbook references~\cite[Section~3.4]{ioannou1996robust} and~\cite[Section~9.6]{khalil} give exponential stability when the slow variation is sufficiently slow. When there are both slow and fast variations, a standard approach is to average out the fast signal $u_f$ to obtain an \emph{average} system (see~\cite[Chapter~V]{hale},~\cite[Chapter~10]{khalil}, or~\cite{sanders}), whose stability can be analyzed under appropriate time-scale separation, and then to recover stability of the original system via perturbation arguments. A number of results have been developed for systems with slow and fast time variations as surveyed in~\cite{Baomtns}. These typically rely on key assumptions, namely whether:
\begin{itemize}[leftmargin=*,nosep]
\item $u_s$ and $u_f$ can be discontinuous in time. Early related works~\cite{Peuteman2002,ChoiShimSeo} restrict to only continuous $u_s$;
\item $u_f$ is periodic or not. Periodicity enables classical averaging techniques as opposed to general averaging methods that are more delicate to analyze;
\item the equilibrium of the original and average systems is common with respect to $u_s$ and $u_f$. If yes (the \emph{common equilibrium} case), exponential stability can be established under suitable conditions. When the equilibrium depends on $u_s$ or $u_f$, one typically obtains practical stability.
\end{itemize}

Note that allowing jumps in $u_s$ makes our framework cover switched systems as an important special case. In the common equilibrium case, asymptotic and exponential stability of that point can be established under arbitrary switching via a common Lyapunov function, or under dwell-time conditions using multiple Lyapunov functions~\cite{hespanha1999,liberzonBook}. When equilibria differ across modes, convergence to a single point is not achievable, and the appropriate notion of stability becomes set-based. In particular, under suitable Lyapunov and dwell-time assumptions, trajectories converge to and remain within a compact set containing the equilibria~\cite{mastellone,alpcan2010stability,dorothy,makarenkov2018dwell}. Under stronger structural conditions, such as the existence of a common quadratic Lyapunov function, one can further guarantee the existence of a compact positively invariant set that is globally attractive even under arbitrary switching~\cite{hafez}, with additional extensions addressing broader system classes and mode configurations~\cite{veer,yinhao}.

Other works addressing systems with multiple time scales include~\cite{teel2003unified} which treats the slow and fast variations separately before coupling them, or~\cite{sanfelice2011,abdelgalil2023multi} (and the references therein) in the context of hybrid systems and singular perturbations. Although hybrid systems may provide an alternative way to analyze system~\eqref{eq:sys}, here we instead treat the slow and fast variations as general exogenous time-varying signals, thereby covering both switching and continuously varying parameters. Returning to slow-fast systems of the form~\eqref{eq:sys}, for linear systems global exponential stability has been studied in a series of works~\cite{ShimLibACM24,ShimLibNAHS25}. For nonlinear systems, semi-global exponential stability has been established under the assumptions that $u_s$ may be discontinuous, the origin is a common equilibrium, i.e., $f(0,\cdot,\cdot)=0$, and the fast signal $u_f$ is periodic~\cite{LibShimCDC24,LibShimTAC25,Baomtns}. When $u_f$ is non-periodic and the equilibrium depends on $u_s$, practical stability results are obtained in~\cite{ChoiShimSeo}, but discontinuities in $u_s$ are not addressed.

This work extends this line of research by considering the \emph{moving equilibrium} case and periodic $u_f$. We establish practical stability in the sense that the state remains within a neighborhood of the moving equilibrium of the average system, which can be arbitrarily reduced by reducing the rate of motion of the equilibrium and accelerating the fast variation. The analysis is illustrated on a nonlinear switched system example with a mode-dependent equilibrium. Compared to our related works, this work is the first to deal with a moving and even jumping equilibrium, naturally covering switched systems with multiple equilibria. Compared with~\cite{ChoiShimSeo}, we bring in new technical tools such as the concept of total variation (see Definition~\ref{def_tv}), capable of capturing both the continuous flows and discrete jumps of $u_s$ as well as the moving equilibrium.

\emph{Notations:} Denote $\RR_{\geq 0} := [0,+\infty)$. Let $|\cdot|$ be the Euclidean vector norm and $\|\cdot\|$ be the induced matrix norm. Note that $\int\|\cdot\|$ has a special meaning as in Definition~\ref{def_tv}. The notion of class-$\K$ function is from~\cite[Definition~4.2]{khalil}.

\section{MAIN RESULT}
Consider system~\eqref{eq:sys} and make the following assumptions.
\begin{assumption}[System~\eqref{eq:sys}]\label{ass_f}
The map $f$ is $C^1$ and its partial derivatives with respect to $x$ and $u_s$ are $C^1$ in $x$. 
\end{assumption}

\begin{assumption}[Fast input]\label{ass_fast}
The function $u_f$ is piecewise continuous and periodic with period $T > 0$. 
\end{assumption}

Thanks to periodicity of $u_f$, the average of $f$, for each fixed $x$ and $u_s$, is defined as
\begin{equation}\label{eq:fav}
\fav(x,u_s) := \frac{1}{T}\int_0^Tf(x,u_s,u_f(s))ds.
\end{equation}
Because $f$ is $C^1$, so is $\fav$. Now, consider the \emph{average} system
\begin{equation}\label{eq:sys_av}
  \dot{x} = \fav(x,u_s).
\end{equation}
We give the class of admissible slow variations.
\begin{assumption}[Slow input~\cite{LibShimTAC25}]\label{ass_slow}
Assume that $u_s$ has finitely many discontinuities
on any bounded interval, is c\`adl\`ag (right-continuous with left limits), is $C^1$ between discontinuities, and Euclidean norm $|\dot{u}_s(\cdot)|$ is Riemann
integrable between discontinuities. The function $u_s(\cdot)$ takes values in a compact and convex set $\Gamma \subset \RR^m$. 
\end{assumption}

We assume that the equilibrium of system~\eqref{eq:sys_av} depends on $u_s$ but is globally exponentially stable when $u_s$ is frozen. This is made formal as follows.

\begin{assumption}[Average system]\label{ass_average}
The average system~\eqref{eq:sys_av} has an isolated equilibrium $\varphi(u_s)$ for each constant input $u_s \in \Gamma$, i.e.,
\begin{equation}
\fav(\varphi(u_s),u_s) = 0, \qquad\forall u_s \in \Gamma,
\end{equation}
such that $\varphi(\cdot)$ is $C^1$ on $\Gamma$. Define:
$$
z := x - \varphi(u_s),
$$
and 
$$
g(z,u_s):=\fav(z + \varphi(u_s),u_s).
$$
There exists a $C^1$ function $V: \RR^n \times \RR^m \to \RR$ such that, for all $(z,u_s) \in \RR^n \times \Gamma$, we have 
\begin{subequations}\label{eq:V_av}
\begin{align}
c_1|z|^2 \leq V(z,u_s) & \leq c_2|z|^2,\label{eq:V_av_12}\\ \frac{\partial V}{\partial z}(z,u_s)g(z,u_s) &\leq -c_3 |z|^2,\label{eq:V_av_3}\\
\left|\frac{\partial V}{\partial u_s}(z,u_s)\right|&\leq c_4 |z|^2,\label{eq:V_av_4}\\
\left|\frac{\partial V}{\partial z}(z,u_s)\right|&\leq c_5 |z|,\label{eq:V_av_5}
\end{align}
\end{subequations}
for some $c_i >0$, $i=1,2,\ldots,5$. 
\end{assumption}
\begin{remark}
Sufficient conditions for such a Lyapunov function are available in the literature. For example, if the equilibrium $z=0$ of the system $\dot z=g(z,u_s)$ is exponentially stable for every fixed $u_s\in\Gamma$, then, under suitable regularity assumptions on $g$, converse Lyapunov theorems guarantee the existence of a function satisfying~\eqref{eq:V_av}; see, e.g.,~\cite[Lemma~9.8]{khalil}. Since our stability result is semi-global, it is sufficient that these properties hold on the compact sets arising in the analysis rather than globally on $\RR^n$. Note also that having global stability ensures that the attraction basins of potentially many disjoint equilibria (for instance, in the case of switched systems with switching equilibria) have a non-empty intersection, avoiding the risk of $x(t)$ being suddenly removed from any attraction basin when $u_s$ jumps.
\end{remark}
Since $\varphi(\cdot)$ is $C^1$ on $\Gamma$ compact, there exists $c_6 > 0$ such that
\begin{equation}\label{eq:c6}
\left|\frac{\partial \varphi}{\partial u_s}(u_s)\right|\leq c_6, \qquad \forall u_s \in \Gamma.
\end{equation}
Because $u_s$ exhibits both continuous- and discrete-time behavior, we define its total variation as follows.
\begin{definition}[Total variation~\cite{gao2018}]\label{def_tv}
Consider a signal $u_s$ on an interval $[t_1,t_2]$ with its discontinuities $d_1<d_2<\ldots< d_m$ such that $t_1 < d_1$ and $d_m \leq t_2$. The \emph{total variation} of $u_s$ on $[t_1,t_2]$ is defined as
\begin{equation}
\int_{t_1}^{t_2}\|du_s\| := \sum_{i=0}^m \int_{d_i}^{d_{i+1}}|\dot{u}_s(t)|dt + \sum_{i=1}^m|u_s(d_i^+) - u_s(d_i^-)|,
\end{equation}
where we set $d_0 := t_1$ and $d_{m+1} := t_2$.
\end{definition}
Note that at the jump times we have $u_s(d^+) = u_s(d) \neq u_s(d^-)$ due to $u_s$ being c\`adl\`ag as in Assumption~\ref{ass_slow}. Our main result, stated next and proven in Section~\ref{sec_proof}, provides practical stability of the moving equilibrium.
\begin{theorem}[Stability of system~\eqref{eq:sys}]\label{theo_stab}
Consider system~\eqref{eq:sys} under Assumptions~\ref{ass_f},~\ref{ass_fast},~\ref{ass_average}, and~\ref{ass_slow}. Assume that there exist $0 \leq \mu < \frac{c_1c_3}{c_2c_4}$ (where the $c_i$'s come from Assumption~\ref{ass_average}) and $\alpha \geq 0$ such that for every interval $[t_1,t_2]$, we have
\begin{equation}\label{eq:cond_tv}
\int_{t_1}^{t_2}\|du_s\| \leq \mu(t_2-t_1)+\alpha.
\end{equation}
Define $\lambda := \frac{1}{2}\left(\frac{c_3}{c_2}-\frac{c_4}{c_1}\mu\right) > 0$. Then, for every $R > 0$, there exist $\e^\star>0$, $\rho >0$, and class-$\K$ functions $\psi_1,\psi_2$ such that, for all $0<\e<\e^\star$, the solution $x(t)$ to system~\eqref{eq:sys} with $|x(0)| \leq R$ satisfies
\begin{multline}\label{eq:xstable}
|x(t)-\varphi(u_s(t))|
\leq \rho e^{-\lambda t}|x(0)-\varphi(u_s(0))| 
\\+ c_6\psi_1(\mu+\alpha) + \psi_2(\mu+\alpha+\e), \qquad \forall t\geq 0.
\end{multline}
\end{theorem}

\begin{figure}[ht]
\centering
\includegraphics[width=0.82\columnwidth]{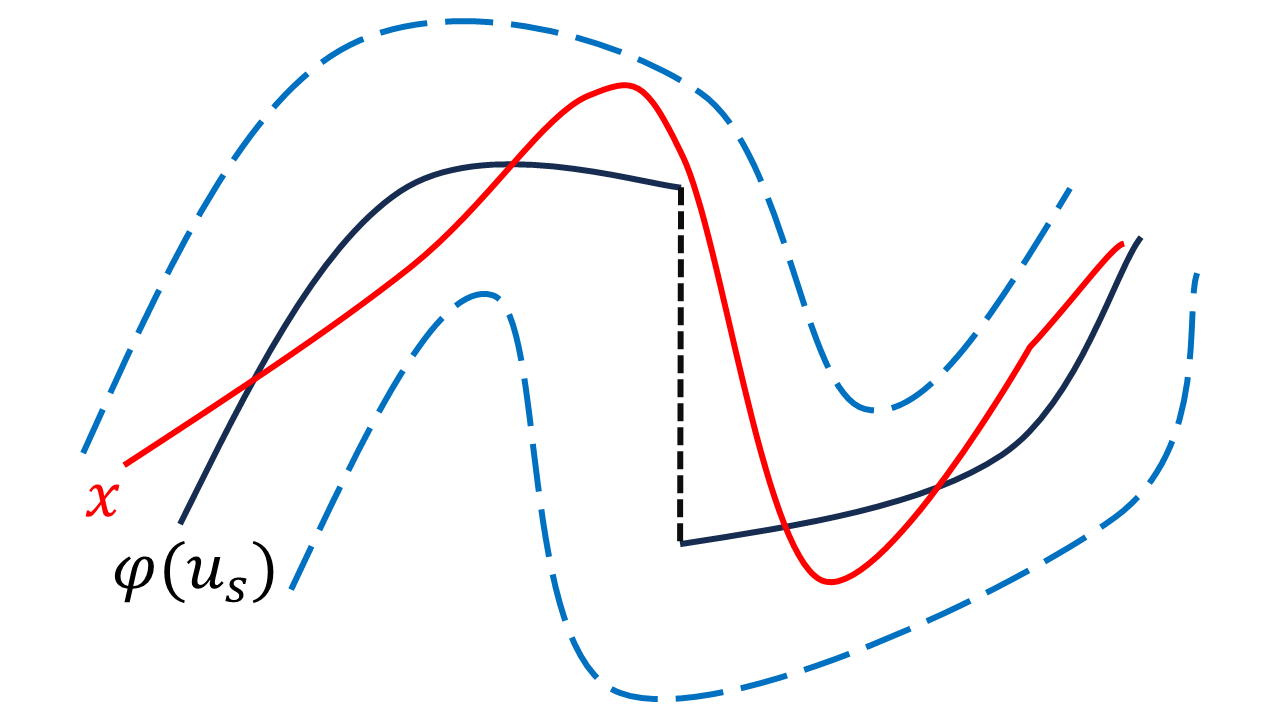}
\caption{Illustration of the practical stability stated by Theorem~\ref{theo_stab}.\label{fig:track}}
\end{figure}

In words, Theorem~\ref{theo_stab} guarantees that if $u_s$ is sufficiently slow (compared to the frozen stability characterized by $c_i$'s---see~\eqref{eq:cond_tv}) and $u_f$ is sufficiently fast, the solution to system~\eqref{eq:sys} remains within a tube around the moving equilibrium $\varphi(u_s)$ of the average system~\eqref{eq:sys_av}, as illustrated in Fig.~\ref{fig:track}. Our stability is \emph{practical} in the sense that the distance between the state and the (moving) equilibrium can be reduced towards $0$ by letting $\mu$, $\alpha$ tend to $0$ (making the equilibrium move less) and similarly for $\e$ (making $u_f$ vary faster, thus improving the approximation by the average system). The achieved stability is intrinsically semi-global due to the dependence of $\e^\star$ on the bounds that appear in the nonlinear analysis, which can only be obtained on compact sets---see Section~\ref{sec_proof}. A global result can be achieved in the linear case, e.g.,~\cite{ShimLibNAHS25}.

\begin{remark}
When $\alpha = 0$, $u_s(t)$ does not jump and satisfies $|\dot{u}_s(t)| \leq \mu$ for all $t \geq 0$, and from~\eqref{eq:xstable} we recover a practical stability result similar to~\cite{ChoiShimSeo}. However, we cannot obtain asymptotic convergence to a stationary equilibrium by letting $\mu=\alpha=0$ in Theorem~\ref{theo_stab}, thus not recovering~\cite{LibShimTAC25,Baomtns}. This difference is because $\varphi(u_s)$ is the equilibrium of the average system~\eqref{eq:sys_av} and \emph{not} of system~\eqref{eq:sys}, i.e., we do not assume that
\begin{equation}
f(\varphi(u_s),u_s,\cdot) = 0,
\end{equation} as in the common equilibrium case. Therefore, Theorem~\ref{theo_stab} should be viewed as an extension of practical stability results for moving equilibria with jumps in $u_s$, rather than as a recovery of the exact exponential stability results available for systems with a common equilibrium. We also see from the term with $c_6$ how the moving equilibrium affects stability. 
\end{remark}

\section{PROOF OF THEOREM~\ref{theo_stab}} \label{sec_proof}
This proof has three main steps: (1) Show some preliminaries and define a change of coordinates $x \mapsto y$ with the resulting $y$ dynamics; (2) Show the boundedness of a Lyapunov function $W(y,u_s)$ in the new coordinates; (3) Deduce the stability in the original coordinates.

\subsection{Preliminaries and Change of Coordinates}
Define the function
\begin{equation}\label{eq:h_def}
h(x,u_s,u_f) := f(x,u_s,u_f) - \fav(x,u_s).
\end{equation}
Then, by definition of $\fav$ and Assumption~\ref{ass_fast}, we have for all $(x,u_s)$
\begin{equation}\label{eq:h_zero_mean}
\int_0^T h(x,u_s,u_f(s))ds = 0.
\end{equation}
Define 
\begin{equation}
w(x,u_s,t) := \int_0^t h(x,u_s,u_f(s))ds.
\end{equation}
Because $u_f$ is periodic and by~\eqref{eq:h_zero_mean}, the function $t\mapsto w(x,u_s,t)$ is $T$-periodic and bounded for each fixed $(x,u_s)$. Since $f$ is $C^1$ and $u_f$ is bounded and periodic, differentiation under the integral sign is valid, and the derivatives of $w$ in $(x,u_s)$ are also $T$-periodic and bounded on compact sets. Hence, for any compact set $D \subset \RR^n \times \Gamma$, there exists a constant $M >0$ such that for all $(x,u_s,t)\in D \times \RR_{\geq 0}$,
\begin{multline}\label{eq:bounds}
|w(x,u_s,t)| \leq M,\\ \left\|\frac{\partial w}{\partial x}(x,u_s,t)\right\| \leq M, \qquad \left\|\frac{\partial w}{\partial u_s}(x,u_s,t)\right\| \leq M.
\end{multline}
Define the Lyapunov function
\begin{equation}
W(z,u_s) := \sqrt{V(z,u_s)},
\end{equation}
where $V$ comes from Assumption~\ref{ass_average}.
We have from~\eqref{eq:V_av_12} that
\begin{equation}\label{eq:sandwichW}
\sqrt{c_1}|z| \leq W(z,u_s) \leq \sqrt{c_2}|z|, \quad \forall (z,u_s) \in \RR^n \times \Gamma.
\end{equation}

Consider a $C^1$ time-varying change of coordinates
\begin{equation}\label{eq:coc}
y := x - \e w(x,u_s(t),t/\e).
\end{equation}
As is standard in averaging theory, we employ a near-identity change of coordinates to eliminate the zero-mean fast oscillations. After this transformation, the dynamics consist of the average system plus perturbation terms of order $O(\e)$, which are amenable to Lyapunov analysis. Reasoning in the same way as~\cite{ChoiShimSeo}, we deduce that there exists $\e^\star_1 > 0$ such that for each $(u_s,t,\e) \in \Gamma \times \RR_{\geq 0} \times (0,\e^\star_1]$, this transformation is $C^1$ and bijective on compact sets of $y$, and it is denoted by $y=\Phi_{t,\e}(x)$ or $x = \Phi_{t,\e}^{-1}(y)$. Then, we deduce that away from the discontinuities of $u_s$,
\begin{subequations}
\begin{multline}\label{eq:ydot}
\dot{y}(t) = \fav(y(t),u_s(t)) + g_0(y(t),u_s(t),t/\e,\e)\dot{u}_s(t) \\+ g_1(y(t),u_s(t),t/\e,\e),
\end{multline}
where $g_0(y,u_s,t/\e,\e) = - \e \frac{\partial w}{\partial u_s}(\Phi_{t,\e}^{-1}(y),u_s,t/\e)$, $g_1(y,u_s,t/\e,\e)= \e F(\Phi_{t,\e}^{-1}(y),y,u_s)w(\Phi_{t,\e}^{-1}(y),u_s,t/\e)- \e \frac{\partial w}{\partial x}(\Phi_{t,\e}^{-1}(y),u_s,t/\e)f(\Phi_{t,\e}^{-1}(y),u_s,u_f(t/\e))$, and
$
F(x,y,u_s) := \int_0^1 \frac{\partial \fav}{\partial x}(\theta x + (1-\theta)y,u_s)d\theta$,
which is continuous because $\fav$ is $C^1$. 
\end{subequations}
Then, it follows from~\eqref{eq:bounds} and the continuity of $F$ 
that there exists $c_g >0$ such that for all $(y,u_s,t,\e) \in D \times \RR_{\geq 0} \times (0,\e^\star_1]$, we have \begin{equation}\label{eq:shim}
|g_0(y,u_s,t/\e,\e)| \leq M \e, \hspace{0.24cm}
|g_1(y,u_s,t/\e,\e)| \leq c_gM\e.
\end{equation}
At a discontinuity $t$ of $u_s$, $y(t)$ experiences a jump (while $x(t)$ does not, or $x(t^+) = x(t^-) = x(t)$) described by
$$
y(t^+) = \Phi_{t^+,\e}(x(t^+)) = \Phi_{t^+,\e}(\Phi_{t^-,\e}^{-1} (y(t^-))).
$$
From the definition of $y$ and continuity of $x(t)$ in $t$, we get
\begin{multline*}
y(t^+) - y(t^-) \\= -\e (w(x(t),u_s(t^+),t/\e) - w(x(t),u_s(t^-),t/\e)).
\end{multline*}
By the Fundamental Theorem of Calculus and~\eqref{eq:bounds}, we get
\begin{equation}\label{eq:ypm}
|y(t^+) - y(t^-)|\leq M \e |u_s(t^+) - u_s(t^-)|.
\end{equation}

\subsection{Boundedness of New Lyapunov Function}
With a slight abuse of notation, we define
\begin{equation}
  V(t):=V(y(t)-\varphi(u_s(t)),u_s(t)),
\end{equation}
and respectively $W(t) = \sqrt{V(t)}$, and study the evolution (flows and jumps) of $V(t)$ and $W(t)$ along $y(t)$ and $u_s(t)$. We first show that for any interval $[t_1,t_2)$ not containing any discontinuity of $u_s$, we have 
\begin{multline}
W(t_2^-) \leq e^{\frac{1}{2}\int_{t_1}^{t_2} \left(-\frac{c_3}{c_2} + \frac{c_4}{c_1} |\dot u_s(s)|\right) ds}W(t_1)
\\+ \int_{t_1}^{t_2} e^{\frac{1}{2}\int_s^{t_2} \left(-\frac{c_3}{c_2} + \frac{c_4}{c_1} |\dot u_s(\tau)|\right) d\tau} \\\times
\left(\frac{c_5(c_6+M\e)}{2\sqrt{c_1}}|\dot{u}_s(s)| + \frac{c_5c_g}{2\sqrt{c_1}}M\e\right) ds.\label{eq:Wdot}
\end{multline}
Indeed, away from the discontinuities of $u_s$, the time derivative of $V$ along~\eqref{eq:ydot} is given by 
\begin{multline*}
\dot{V}(t)= \frac{\partial V}{\partial y}(y(t)-\varphi(u_s(t)),u_s(t))(\fav(y(t),u_s(t)) \\+ g_0(y(t),u_s(t),t/\e,\e)\dot{u}_s(t) + g_1(y(t),u_s(t),t/\e,\e)) \\
+ \bigg(\frac{\partial V}{\partial u_s}(y(t)-\varphi(u_s(t)),u_s(t)) \\- \frac{\partial V}{\partial y}(y(t)-\varphi(u_s(t)),u_s(t))\frac{\partial \varphi}{\partial u_s}(u_s(t))\bigg)\dot{u}_s(t).
\end{multline*}
We get from Assumption~\ref{ass_average} and~\eqref{eq:shim} that on $D$,
\begin{align*}
\dot{V}(t) &\leq \left(-\frac{c_3}{c_2} +\frac{c_4}{c_1}|\dot{u}_s(t)|\right) V(t) \\&\qquad{}+ \frac{c_5(c_6+M\e)}{\sqrt{c_1}}\sqrt{V(t)}|\dot{u}_s(t)| + \frac{c_5c_g}{\sqrt{c_1}}M\e\sqrt{V(t)}.
\end{align*}
When $V(t) \neq 0$, dividing both sides by $2\sqrt{V(t)}$ and noting that $\dot{W}(t) = \frac{\dot{V}(t)}{2\sqrt{V(t)}}$, we get
\begin{multline*}
\dot{W}(t) \leq \frac{1}{2}\left(-\frac{c_3}{c_2} +\frac{c_4}{c_1}|\dot{u}_s(t)|\right) W(t)\\+ \frac{c_5(c_6+M\e)}{2\sqrt{c_1}}|\dot{u}_s(t)| + \frac{c_5c_g}{2\sqrt{c_1}}M\e.
\end{multline*}
Using the standard comparison principle with this inequality and some reasoning using the continuity of $W$ for the case $V(t) = 0$, we obtain~\eqref{eq:Wdot}. At a discontinuity $t$ of $u_s$, we have $y(t^+) = \Phi_{t^+,\e}(x(t))$ and $u_s(t^+) \neq u_s(t^-)$. We now show that at each discontinuity of $u_s$, the jump of $W$ satisfies
\begin{multline}
W(t^+) \leq e^{\frac{c_4}{2c_1}|u_s(t^+) - u_s(t^-)|}W(t^-) 
\\+ \frac{c_5}{2\sqrt{c_1}}(M \e + c_6)|u_s(t^+) - u_s(t^-)|.\label{eq:Wplus}
\end{multline}
To do that, we decompose
\begin{align}\label{eq:Wsplit}
W(t^+) - W(t^-)&= W(y(t^+)-\varphi(u_s(t^+)),u_s(t^+))\notag\\&{} - W(y(t^-)-\varphi(u_s(t^-)),u_s(t^+))
 \notag\\&{}+ W(y(t^-)-\varphi(u_s(t^-)),u_s(t^+)) \notag\\&{}- W(y(t^-)-\varphi(u_s(t^-)),u_s(t^-)).
\end{align}
Note that the non-differentiability of $W$ at the equilibrium prevents us from applying the Mean Value Theorem to this function. One way to deal with this issue is by \emph{regularizing} $W$, defining $W_\delta(t) := \sqrt{V(t) + \delta}$ for $\delta > 0$, which is $C^1$. For the first difference on the right-hand side of~\eqref{eq:Wsplit}, by the Mean Value Theorem in the first argument of $W_\delta$, we deduce that there exists $\overline{z} = r (y(t^+)-\varphi(u_s(t^+))) + (1 - r)(y(t^-)-\varphi(u_s(t^-)))$ for some $r \in [0,1]$ such that
\begingroup
\small
\begin{multline*}
W_\delta(y(t^+)-\varphi(u_s(t^+)),u_s(t^+))
- W_\delta(y(t^-)-\varphi(u_s(t^-)),u_s(t^+)) \\
\leq \left|\frac{\partial W_\delta}{\partial z}
(\overline{z},u_s(t^+))\right|
\left|y(t^+) - \varphi(u_s(t^+))
- (y(t^-)- \varphi(u_s(t^-)))\right|.
\end{multline*}
\endgroup
\noindent
From $W_\delta=\sqrt{V+\delta}$ and~\eqref{eq:V_av}, we have
\begin{align*}
\left|\frac{\partial W_\delta}{\partial z}(\overline{z},u_s(t^+))\right|
&= \frac{1}{2\sqrt{V(\overline{z},u_s(t^+))+\delta}}\left|\frac{\partial V}{\partial z}(\overline{z},u_s(t^+))\right|\\
&\leq \frac{c_5|\overline{z}|}{2\sqrt{c_1|\overline{z}|^2+\delta}}\leq \frac{c_5}{2\sqrt{c_1}}.
\end{align*}
Note that this inequality is true even if $|\overline{z}| = 0$. Hence, combining this with the triangle inequality and taking the limit as $\delta \to 0$ (using the fact that the bound on the right-hand side is independent of $\delta$), we have
\begingroup
\small
\begin{multline}
W(y(t^+)-\varphi(u_s(t^+)),u_s(t^+))
- W(y(t^-)-\varphi(u_s(t^-)),u_s(t^+)) \\
\leq \frac{c_5}{2\sqrt{c_1}}
(|y(t^+) - y(t^-)|+
|\varphi(u_s(t^+)) - \varphi(u_s(t^-))|).
\end{multline}
\endgroup
\noindent
Using~\eqref{eq:ypm} and~\eqref{eq:c6}, we have
\begin{multline}\label{eq:jump_state}
W(y(t^+)-\varphi(u_s(t^+)),u_s(t^+))\\ - W(y(t^-)-\varphi(u_s(t^-)),u_s(t^+)) 
\\\leq \frac{c_5 }{2\sqrt{c_1}}(M \e+c_6)|u_s(t^+) - u_s(t^-)|.
\end{multline}
For the second difference of~\eqref{eq:Wsplit}, by the Mean Value Theorem in the second argument of $W_\delta$, we get that there exists $\overline{u}_s = ru_s(t^+) + (1-r)u_s(t^-)$ for some $r \in [0,1]$, which is in $\Gamma$ thanks to convexity, such that
\begingroup
\small
\begin{multline}\label{eq:mmp}
W_\delta(y(t^-)-\varphi(u_s(t^-)),u_s(t^+)) \\
- W_\delta(y(t^-)-\varphi(u_s(t^-)),u_s(t^-))\\
\leq \left|\frac{\partial W_\delta}{\partial u_s}
(y(t^-)-\varphi(u_s(t^-)), \overline{u}_s)\right|
|u_s(t^+) - u_s(t^-)|.
\end{multline}
\endgroup
\noindent
Using~\eqref{eq:V_av} and~\eqref{eq:sandwichW}, we get
\begin{align*}
&\left|\frac{\partial W_\delta}{\partial u_s}(y(t^-)-\varphi(u_s(t^-)), \overline{u}_s)\right|\\
&= \scalebox{0.99}{$\frac{1}{2\sqrt{V(y(t^-)-\varphi(u_s(t^-)), \overline{u}_s) +\delta}}\left|\frac{\partial V}{\partial u_s}(y(t^-)-\varphi(u_s(t^-)), \overline{u}_s)\right|$}\\
&\leq \frac{1}{2\sqrt{c_1|y(t^-)-\varphi(u_s(t^-))|^2 + \delta}} c_4 |y(t^-)-\varphi(u_s(t^-))|^2\\
&\leq \frac{c_4}{2\sqrt{c_1}} |y(t^-)-\varphi(u_s(t^-))|\\&\leq\frac{c_4}{2c_1}W(y(t^-)-\varphi(u_s(t^-)),u_s(t^-)).
\end{align*}
Note that this inequality is true even if $|y(t^-)-\varphi(u_s(t^-))| = 0$. Therefore, combining~\eqref{eq:mmp} with the inequality $1+z \leq e^z$ for $z \geq 0$ and taking the limit as $\delta \to 0$ (still noting that the bound on the right-hand side is independent of $\delta$), we have
\begin{multline}
W(y(t^-)-\varphi(u_s(t^-)),u_s(t^+))
\\\hspace{-0.1cm}\leq e^{\frac{c_4}{2c_1}|u_s(t^+) - u_s(t^-)|}W(y(t^-)-\varphi(u_s(t^-)),u_s(t^-)).\label{eq:jump_input}
\end{multline}
Combining~\eqref{eq:jump_state} and~\eqref{eq:jump_input}, we get~\eqref{eq:Wplus}. After bounding the flow and jump dynamics of $W(t)$, we now bound its value. To simplify notation, we introduce the constants: $
c_7 := \frac{c_5 c_6}{2\sqrt{c_1}}$, $ 
c_8 := \frac{c_5 M}{2\sqrt{c_1}}$, $c_9 := \frac{c_5 c_g M}{2\sqrt{c_1}}$. Define $
a(t):=-\frac{c_3}{2c_2}+\frac{c_4}{2c_1}|\dot{u}_s(t)|$, $b(t):=(c_7+c_8\e)|\dot{u}_s(t)|+c_9\e$, for all $t$ outside the discontinuities of $u_s$, and for each discontinuity $d_i$ of $u_s$, define $\Delta_i:=\frac{c_4}{2c_1}|u_s(d_i^+)-u_s(d_i^-)|$, $\delta_i:=(c_7+c_8\e)|u_s(d_i^+)-u_s(d_i^-)|$. Fix $t\geq 0$, and let $0=d_0<d_1<\cdots<d_m<d_{m+1}=t$ be the discontinuities of $u_s$ in $[0,t]$. Then, we get
\begin{subequations}
\begin{align}
W(d_{i+1}^-)&\leq e^{\int_{d_i}^{d_{i+1}}a(\tau)d\tau}W(d_i^+)\notag\\&{}+\int_{d_i}^{d_{i+1}}e^{\int_s^{d_{i+1}}a(\tau)d\tau}b(s)ds,\quad i=0,1,\dots,m,\label{eq:flow_compact}\\
W(d_i^+)&\leq e^{\Delta_i}W(d_i^-)+\delta_i,\qquad i=1,2,\dots,m.\label{eq:jump_compact}
\end{align}
\end{subequations}
Iterating~\eqref{eq:flow_compact}--\eqref{eq:jump_compact} over $[0,t]$, we obtain
\begin{multline}
W(t)\leq\exp\left(\sum_{i=0}^m\int_{d_i}^{d_{i+1}}a(\tau)d\tau+\sum_{i=1}^m\Delta_i\right)W(0)\\
+\sum_{i=0}^m\int_{d_i}^{d_{i+1}}\exp\bigg(\int_s^{d_{i+1}}a(\tau)d\tau\\+\sum_{j=i+1}^m\int_{d_j}^{d_{j+1}}a(\tau)d\tau+\sum_{j=i+1}^m\Delta_j\bigg)b(s)ds\\
+\sum_{i=1}^m\exp\left(\sum_{j=i}^m\int_{d_j}^{d_{j+1}}a(\tau)d\tau+\sum_{j=i+1}^m\Delta_j\right)\delta_i.\label{eq:W_iter}
\end{multline}
We can then show that for any $t\geq 0$ (lengthy proof omitted),
\begin{multline}
W(t)\leq e^{-\lambda t+\frac{c_4}{2c_1}\alpha}W(0)\\+e^{\frac{c_4}{2c_1}\alpha}\left(\frac{c_9}{\lambda}\e+2(c_7+c_8\e)\left(\alpha+\frac{\mu}{\lambda}\right)\right),
\end{multline}
with $\lambda$ defined in Theorem~\ref{theo_stab}. From~\eqref{eq:sandwichW}, we get that for any $t\geq 0$,
\begin{multline}
|y(t) - \varphi(u_s(t))| \leq \sqrt{\frac{c_2}{c_1}}e^{-\lambda t+\frac{c_4}{2c_1}\alpha} |y(0) - \varphi(u_s(0))|\\+ \frac{e^{\frac{c_4}{2c_1}\alpha}}{\sqrt{c_1}}\left(\frac{c_9}{\lambda}\e
+ 2(c_7 + c_8\e)\left(\alpha + \frac{\mu}{\lambda}\right)\right).
\end{multline}

\subsection{Stability of System~\eqref{eq:sys}}
From~\eqref{eq:coc}, we have $x(0) = y(0)$. From~\eqref{eq:coc} and~\eqref{eq:bounds}, 
we have for all $t \geq 0$,
\begin{equation}\label{eq:xy_relation}
|x(t) - y(t)| = \e |w(x(t),u_s(t),t/\e)| \leq M\e.
\end{equation} 
Using the triangle inequality together with~\eqref{eq:xy_relation}, the previous bound on $|y(t) - \varphi(u_s(t))|$, and that $x(0) = y(0)$, 
we obtain for all $t \geq 0$,
\begin{multline}
|x(t) - \varphi(u_s(t))|\leq M\e + \sqrt{\frac{c_2}{c_1}}e^{-\lambda t+\frac{c_4}{2c_1}\alpha}|x(0) - \varphi(u_s(0))|\\+ \frac{e^{\frac{c_4}{2c_1}\alpha}}{\sqrt{c_1}}\left(\frac{c_9}{\lambda}\e
+ 2(c_7 + c_8\e)\left(\alpha + \frac{\mu}{\lambda}\right)\right).\label{eq:x_bound}
\end{multline}
Fix $R > 0$ and consider initial conditions satisfying $|x(0)| \leq R$. 
Recall that $
z(t) := x(t) - \varphi(u_s(t))$ and $z(0) = x(0) - \varphi(u_s(0))$. Using boundedness of $\varphi$ on $\Gamma$, we deduce that there exists $c_{10} > 0$ such that $|\varphi(u_s)| \leq c_{10}$ for all $u_s \in \Gamma$, and therefore $|x| \leq |z| + c_{10}$. Recall that $c_7$ is linear in $c_6$. Hence, there exists a class-$\K$ function $\psi_1$ such that
$
2\frac{e^{\frac{c_4}{2c_1}\alpha}}{\sqrt{c_1}}c_7
\left(\alpha+\frac{\mu}{\lambda}\right)
\leq
c_6\psi_1(\mu+\alpha)$.
Moreover, all the remaining terms in~\eqref{eq:x_bound} that do not multiply $c_6$
are non-negative, vanish when $\mu=\alpha=\e=0$, and their sum can be upper bounded by a class-$\K$ function $\psi_2$ of $\mu+\alpha+\e$, i.e., $M\e
+
\frac{e^{\frac{c_4}{2c_1}\alpha}}{\sqrt{c_1}}
\left(
\frac{c_9}{\lambda}\e
+
2c_8\e
\left(\alpha+\frac{\mu}{\lambda}\right)
\right)
\leq
\psi_2(\mu+\alpha+\e)$. Note that $\lambda$ defined in Theorem~\ref{theo_stab} does depend on $\mu$, but it increases as $\mu$ decreases, so the bounding still holds.
Consequently, from~\eqref{eq:x_bound}, we have for all $t \geq 0$ 
that $|x(t)| \leq \rho e^{-\lambda t} |z(0)| + c_6\psi_1(\mu+\alpha) + \psi_2(\mu+\alpha+\e) + c_{10}$,
where $\rho := \sqrt{\frac{c_2}{c_1}}e^{\frac{c_4}{2c_1}\alpha}$. Choose $R^\prime > 0$ large enough so that
$
\rho |z(0)| + c_6\psi_1(\mu+\alpha) + \psi_2(\mu+\alpha+\e) + c_{10} \leq R^\prime$
for all $|x(0)| \leq R$ and sufficiently small $\e$. Then $x(t) \in \Omega := \{ x \in \RR^n : |x| \leq R^\prime\}$, which is compact, for all $t \geq 0$, which together with the compactness of $\Gamma$ ensures that all previous constant estimates (in particular~\eqref{eq:bounds}) remain valid along the trajectory. Therefore, the above bounds in $(x,u_s)$ hold in time, and Theorem~\ref{theo_stab} follows.

\section{ILLUSTRATION: SWITCHED SYSTEM}
We consider a two-dimensional switched system with nonlinear dynamics
\begin{equation}\label{eq:sys_switch}
\dot x(t) = \frac{1}{1+|x(t)|^2}A_{\sigma(t)}(x(t) - x_{\sigma(t)}^\star),
\end{equation}
and four modes, where $\sigma(t)\in\{1,2,3,4\}$ is a slow-fast switching signal defined later. The first two modes have the same equilibrium $x^\star_1 = x^\star_2 = (-1,0) =: x^\star_{av,1}$, while the other two have another equilibrium $x^\star_3 = x^\star_4 = (1,0) =: x^\star_{av,2}$. The mode dynamics matrices $A_i$ ($i = 1,2,3,4$) are constructed so that
\begin{align*}
\frac{1}{2}(A_1 + A_2) & = \begin{pmatrix}
-0.3 & -20/3 \\
0.6  & -0.3
\end{pmatrix} =: A_{av,1},\\
\frac{1}{2}(A_3 + A_4) &= \begin{pmatrix}
-0.3 & -0.6 \\
20/3 & -0.3
\end{pmatrix} =: A_{av,2}.
\end{align*}
Moreover, we choose some of these $A_i$'s to have eigenvalues with positive real parts, so that system~\eqref{eq:sys_switch} frozen at such modes is unstable.

The switching law is designed to be the composition of slow switching between the two groups $\{1,2\}$ and $\{3,4\}$ and a fast switching within each group, both of which are periodic. More formally, let $\e_s>0$ and $\e_f>0$ be tuning parameters for the slow and fast time scales (typically $\e_s \gg \e_f$), respectively, and define the periodic reference signals
$$
  r_s(t) = \sin(t/\e_s),\qquad
  r_f(t) = \sin(t/\e_f).
$$
With these, the slow group selector and the fast within-group selector (both periodic) are
$$
  q_s(t)=
\begin{cases}
1, & r_s(t)\geq 0,\\
2, & \text{otherwise},
\end{cases} \quad q_f(t)=
\begin{cases}
1, & r_f(t)\geq 0,\\
2, & \text{otherwise}.
\end{cases}
$$
Finally, the mode $\sigma(t)\in\{1,2,3,4\}$ is defined by
\begin{equation}
\sigma(t)=
\begin{cases}
1, & q_s(t)=1,\ q_f(t)=1,\\
2, & q_s(t)=1,\ q_f(t)=2,\\
3, & q_s(t)=2,\ q_f(t)=1,\\
4, & q_s(t)=2,\ q_f(t)=2.
\end{cases}
\end{equation}
For analysis purposes, it is relevant to introduce the
following switched system with two modes
\begin{equation}\label{eq:sys_switch_ave}
\dot{x} = \frac{1}{1+|x|^2}A_{av,1}(x - x^\star_{av,1}), \dot{x} = \frac{1}{1+|x|^2}A_{av,2}(x - x^\star_{av,2}),
\end{equation}
which is exponentially stable when frozen at each mode, as can be shown with the quadratic Lyapunov function $V_i(x) = (x - x^\star_{av,i})^\top P_i (x - x^\star_{av,i})$ where each $P_i = P_i^\top > 0$ solves $A_{av,i}^\top P_i + P_i A_{av,i} = -I$ for $i = 1,2$,
which exists because each $A_{av,i}$ is Hurwitz. However, there is no common solution $P = P^\top > 0$ for both $i = 1,2$.

With our construction of system~\eqref{eq:sys_switch}, when $\e_f$ is sufficiently small, the fast switching within each group yields average dynamics that are close to system~\eqref{eq:sys_switch_ave}, i.e., the dynamics matrix is $A_{av,1}$ and the equilibrium is $x^\star_{av,1}$ on intervals where $q_s(t)=1$, and $A_{av,2},x^\star_{av,2}$ on intervals where $q_s(t)=2$, while the slow switching between these two average dynamics is tuned by $\e_s$. Moreover, there is a mode-dependent quadratic Lyapunov function $V_i$ for the average system~\eqref{eq:sys_switch_ave}, guaranteeing exponential stability when this system is frozen at each mode, so that Assumption~\ref{ass_average} is satisfied. Next, we will simulate system~\eqref{eq:sys_switch} with different choices of $\e_s$, $\e_f$, and plot the trajectories in the state space. 

\begin{figure}[H]
\centering
\includegraphics[width=0.324\columnwidth,height=0.246\columnwidth]{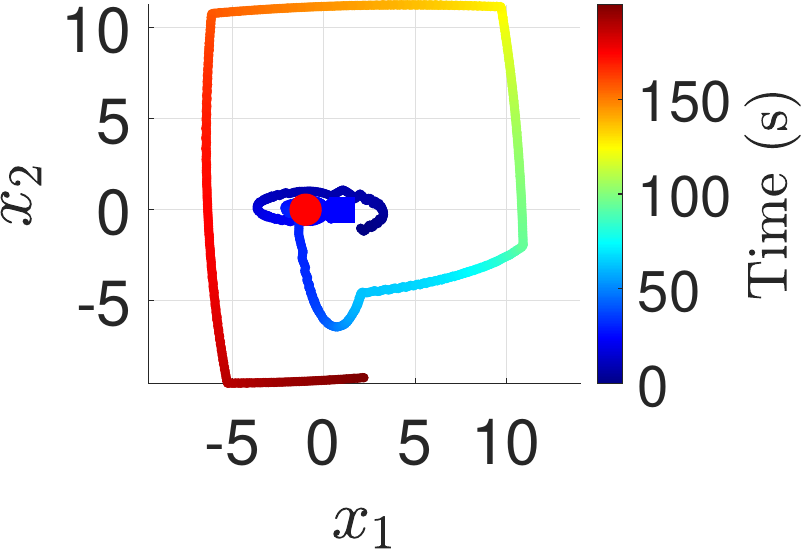}
\includegraphics[width=0.324\columnwidth,height=0.246\columnwidth]{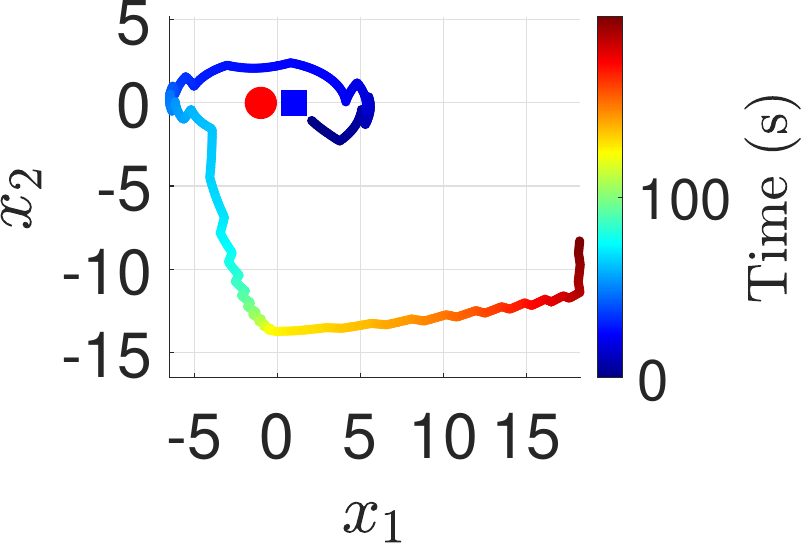}
\includegraphics[width=0.324\columnwidth,height=0.246\columnwidth]{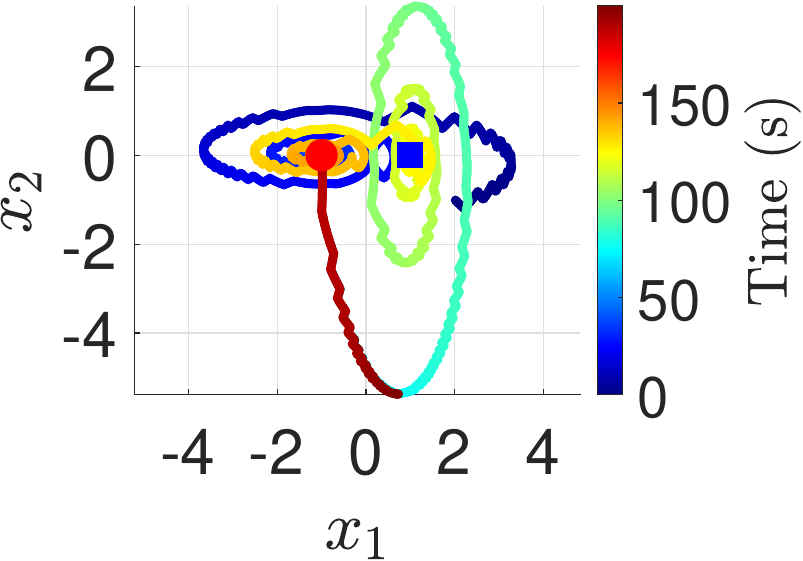}
\caption{Trajectories of system~\eqref{eq:sys_switch} in the state space for respectively: $\e_s = 10$, $\e_f = 0.1$ (left); $\e_s = 20$, $\e_f = 1$ (middle); $\e_s = 20$, $\e_f = 0.1$ (right). The dot and square are the equilibria. The color indicates time.\label{fig1}}
\end{figure}

Simulation results with $\e_s = 10$, $\e_f = 0.1$ and with $\e_s = 20$, $\e_f = 1$, shown in Figure~\ref{fig1}, indicate that the switched system~\eqref{eq:sys_switch} is unstable. In the first case, the slow switching is not slow enough, and so the (frozen) stability of $A_{av,1},A_{av,2}$ does not carry over. In the second case, the fast switching is not fast enough, and so the averaging cannot make system~\eqref{eq:sys_switch} inherit stability from the average dynamics~\eqref{eq:sys_switch_ave}. We then experiment with $\e_s = 
20$ and $\e_f = 0.1$, which are respectively slow and fast enough, and practical stability is achieved as shown in Figure~\ref{fig1}. The state keeps getting attracted towards the current equilibrium and remains in a neighborhood containing the equilibria. This illustrates the result in Theorem~\ref{theo_stab}, which in this case is consistent with the ones for switched systems with switching equilibria such as~\cite{mastellone,alpcan2010stability,makarenkov2018dwell}. Moreover, note that our framework is not limited to switched systems but applies to a much broader class of nonlinear time-varying systems.

\section{CONCLUSION}
We study the stability of nonlinear systems with a slowly moving equilibrium and periodic fast variation. We establish a practical stability result ensuring that the state remains within a neighborhood of the moving equilibrium of the average system. The analysis allows discontinuities in the time variations, covering switched systems as a special case.

Future work will consider the even more general setting of a moving equilibrium and \emph{non-periodic} fast variations, which requires general averaging techniques. Some challenges compared to this periodic case are:
\begin{itemize}[leftmargin=*,nosep]
\item The average map $\fav$ obtained from general averaging, unlike that in~\eqref{eq:fav}, may not be $C^1$ even if $f$ is. Related sufficient conditions should be established;
\item The transformation $w$ and its derivatives may not be bounded as easily as~\eqref{eq:bounds}. In fact, another $w$ will be used.
\end{itemize}
We will also investigate related applications such as HIV treatment~\cite{ChoiShimSeo} or PWM-based control~\cite{pwm2020}.

\bibliographystyle{IEEEtran}
\bibliography{ref}
\end{document}